\documentclass{article}
\usepackage{graphics}

\begin{document}

\begin{center}
{\Large The Effects of Moore's Law and Slacking
\footnote{This paper took 2 days to write} on Large Computations}\\
\smallskip
{Chris Gottbrath, Jeremy Bailin, Casey Meakin, Todd Thompson, J.J.~Charfman}\\
Steward Observatory, University of Arizona
\end{center}

\medskip

\begin{abstract}

We show that, in the context of Moore's Law, overall productivity can be 
increased for large enough computations by `slacking' or waiting
for some period of time before purchasing a computer and beginning
the calculation. 

\end{abstract}

According to Moore's Law, the computational power available at a particular
price doubles every 18~months. Therefore it is conceivable that for
sufficiently large numerical calculations and fixed budgets, computing power 
will improve 
quickly enough that the calculation will finish faster if we wait until
the available computing power is sufficiently better and start the
calculation then.

\begin{figure}[h]\caption{\label{overall}}
\scalebox{0.45}{\rotatebox{270}{\includegraphics{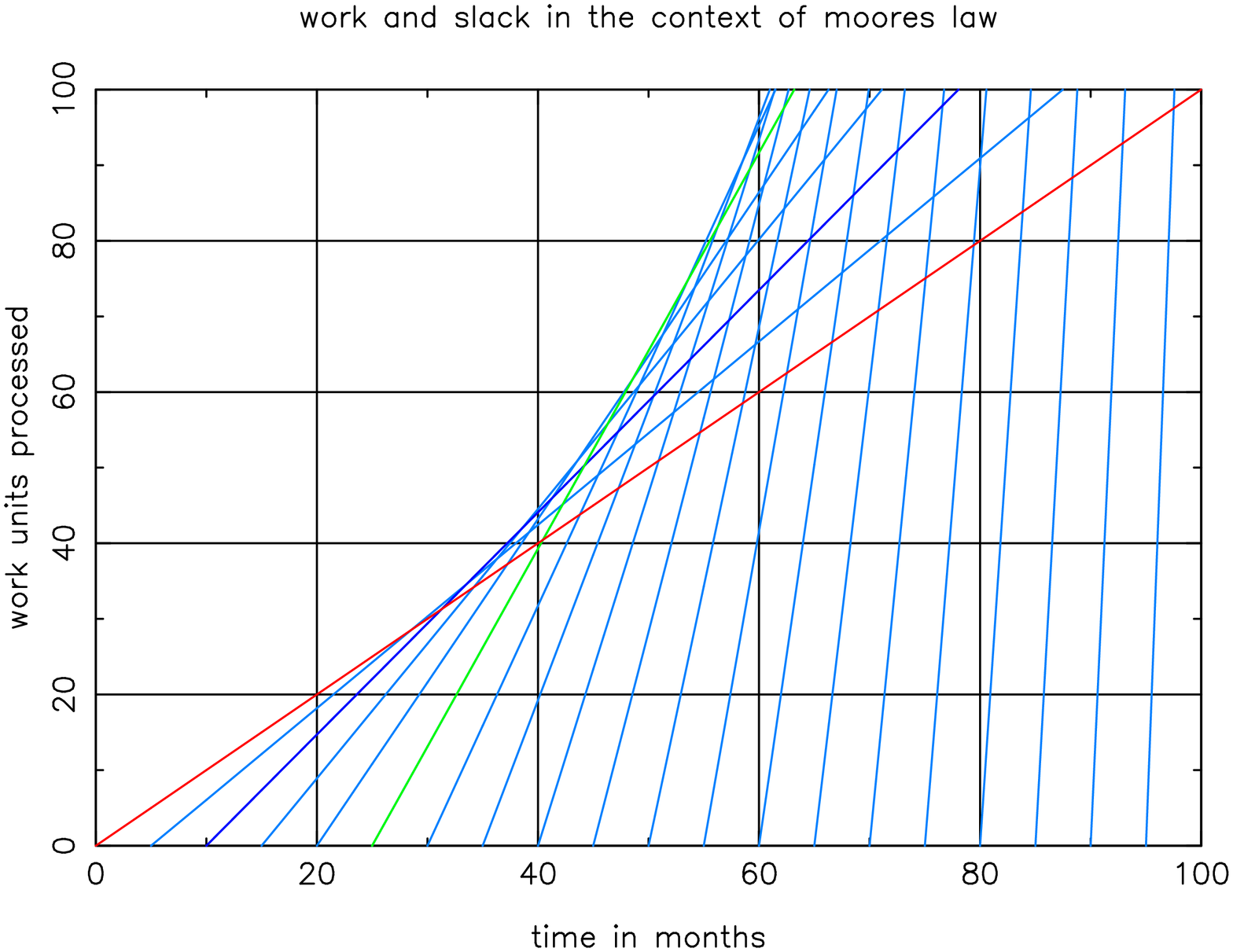}}}
\end{figure}

\vspace{5mm}

This is illustrated in the above plot. Work is measured in units of
whatever a current machine can accomplish in one month
and time is measured in 
months. The red line denotes the amount of work completed if you start
calculating now. However, according to Moore's law the speed of computation,
$R$, grows as
$R_t=R_0 2^{t/18 mo}$. In our units $R_0=1$. If you wait some 
amount of 
time, then buy a new computer and begin the computation,
Moore's law ensures that the new computer will be faster, and you
will get a steeper performance curve. The blue lines illustrate the performance
you will get if you wait five, ten, or more months.
This begins to pay off if the calculation 
is large enough. For example, by looking at the green line we see
that waiting for 25 months pays off for any 
calculation larger than 40 work units;
you could start a computation now, calculate for 40
months, and get a certain amount of work done. Alternately, you could
go to the beach for 2 years, then come back and buy a new computer and 
compute for a year, and get the same amount of work done.

Specifically, let $N$ be the amount of work involved in the 
calculation, which in our notation is the number of months the
calculation would take with current hardware, $R_1$ be the rate of operations
at some future time, and $t_1=N/R_1$ be the time the calculation takes at
that future time.
If we wait a ``slack time'' $s$, then begin 
calculating with the newer faster computer we will finish at 
\begin{equation}t_{finish} = t_1 + s \label{t_finish}\end{equation}
ie. the total time it takes is the time of the computation in the future
plus how 
long we slack. Finally, if all the times are measured in months, then 
Moore's Law tells us that the rate increases exponentially:
\begin{equation} R_1 = 2^{s/18}. \label{rate equation} \end{equation}

We now calculate how long we can slack and still get the same amount done as 
if we had started immediately. $N$ is the time it would take
for the calculation to complete if started now, so from equations 
\ref{t_finish} and \ref{rate equation} we have 
\begin{equation} t_{finish} = N = s + t_1 = s + N (2^{-{s/18 mo}})
	\label{tfinish} \end{equation}
$$ N = 2^{s/18} (N - s) $$
$$ N (2^{s/18} -1) = s (2^{s/18}) $$
$$ N = \frac{s (2^{s/18})} {2^{s/18}-1} $$
\begin{equation} N = \frac{s}{1 - 2^{-s/18}} \end{equation}

This shows the relationship between the total work and slack. Note from 
looking at figure \ref{overall} that this is also the largest amount 
of slacking that can be done and still get the work done in N months.

Note that the size of the calculation does not vanish as $s\rightarrow 0$,
ie. there is a minimum calculation for which it is ever worth it to
slack. This is reassuring, since otherwise it would \emph{always} be
worth it to wait and we would never get anything done. $t_0(s)$
is undefined at $s=0$, but taking the limit using L'H\^opital's
rule,
\begin{equation} \lim_{s\rightarrow 0} t_0 = \frac{1}
	{(\frac{\ln 2}{18}) 2^{-s/18}} = \frac{18}{\ln 2} = t_c
	\approx 26.0 \end{equation}
Therefore, any calculation that currently takes less than 26 months will
finish earliest if started immediately. We define this to be the 
critical timescale $t_c$ which is the e-folding time of Moore's law.

If we define the productivity as the work divided by time, we can see
how much our productivity improves as a result of our slacking. For a
calculation of a given size, we define the productivity enhancement
factor $P$ to be the ratio of the time it takes to finish the job now
to the time it would take to finish the job after slacking for
a time $s$.
$$ P = \frac{N}{t_{finish}} = \frac{N} {s + N (2^{-s/18})} $$
\begin{equation} P = \left[ {s\over N} + 2^{-s/18} \right]^{-1} \end{equation}
The surface of the productivity enhancement in the $N$-$s$ plane is shown
below:

\begin{figure}[h]\caption{\label{pef}}
\input{pef.pstex}
\end{figure}

Even better, you will notice in figure \ref{overall} that the dark blue line (denoting slacking for 
ten months) passes the 40 work unit mark 5 months ahead of the 
red line. This suggests that by fine tuning your slacktitude you can actually 
accomplish 
more than either the lazy bum at the beach for two years or the hard working 
sucker who got started immediately. Indeed with a little bit of algebra we 
convince ourselves that there exists an optimal slack time $s\star$. 

We start by noting that the time before finishing a job $t_{finish}$,
as given in equation \ref{tfinish},
goes through a minimum at $s=s\star$.
We set the derivative of $t_{finish}$ with respect to slack equal to zero and
solve for $s\star$. 
\begin{equation}
s\star=\frac{18}{\ln{2}} \ln{\left( \frac{\ln{2}}{18} N\right)} =
t_c \ln{\left(  \frac{N}{t_c}\right)} \label{sstar}
\end{equation}

\begin{figure}[h]\caption{\label{optimal}}
\scalebox{0.45}{\rotatebox{270}{\includegraphics{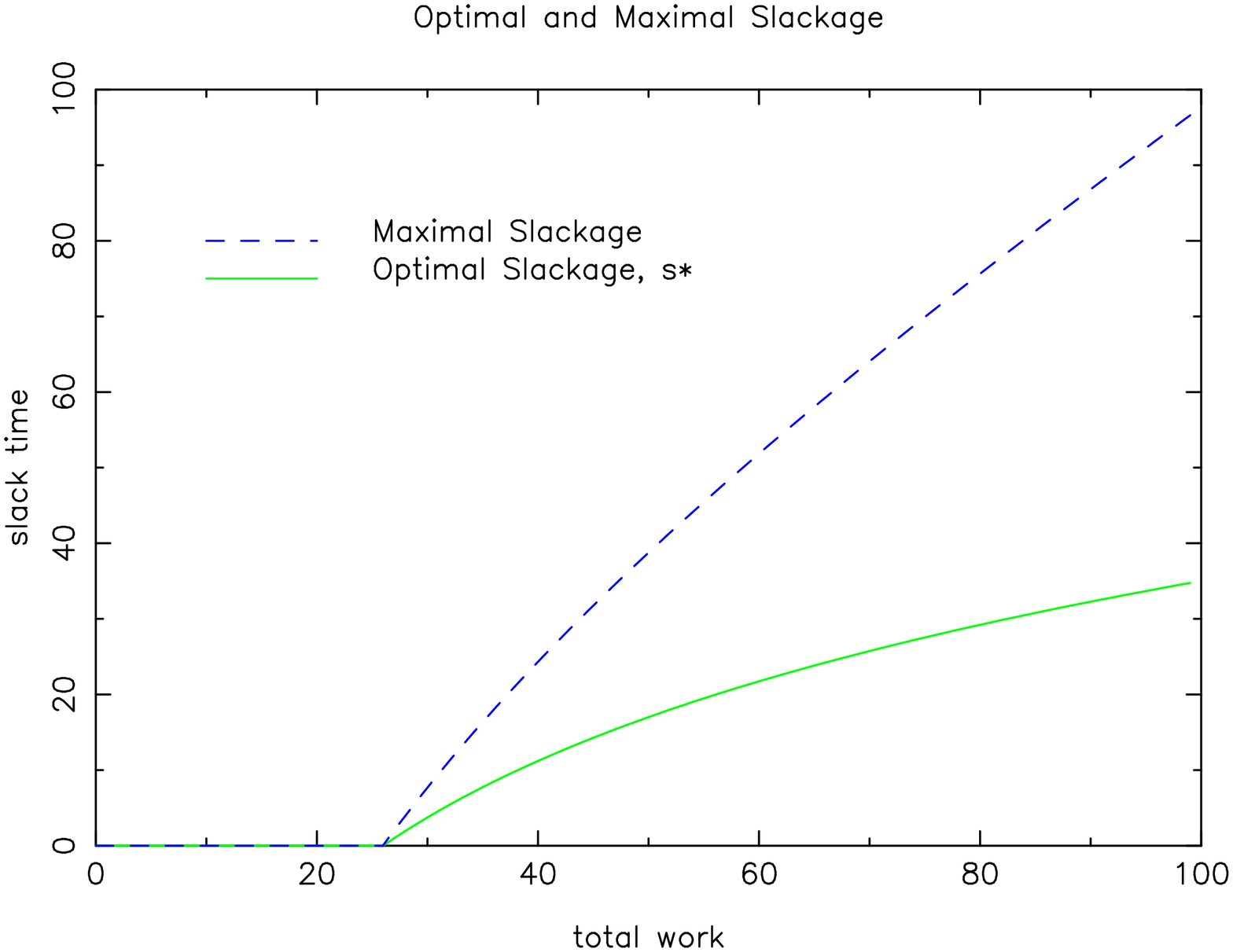}}}
\end{figure}

\vspace{5mm}

This means that if we want to slack for a year, we should choose a task that 
would normally take 41.2 months to complete at current processor speeds. 
After our optimal year of goofing around we buy a new computer,
put our noses to the grindstone,
and finish the calculation $t_c$ months later, having saved ourselves 
3.25 months worth of total time (plus having been able to slack for a year and 
honestly call it productive).

The time to optimally finish the task is simply given by substituting
equation \ref{sstar} into equation \ref{tfinish}:
\begin{equation} t\star = s\!\star + N (2^{-s\star/18})
	 = t_c \left[1 + \ln(N/t_c)\right] \end{equation}

\begin{figure}[h]\caption{\label{besttime}}
\scalebox{0.45}{\rotatebox{270}{\includegraphics{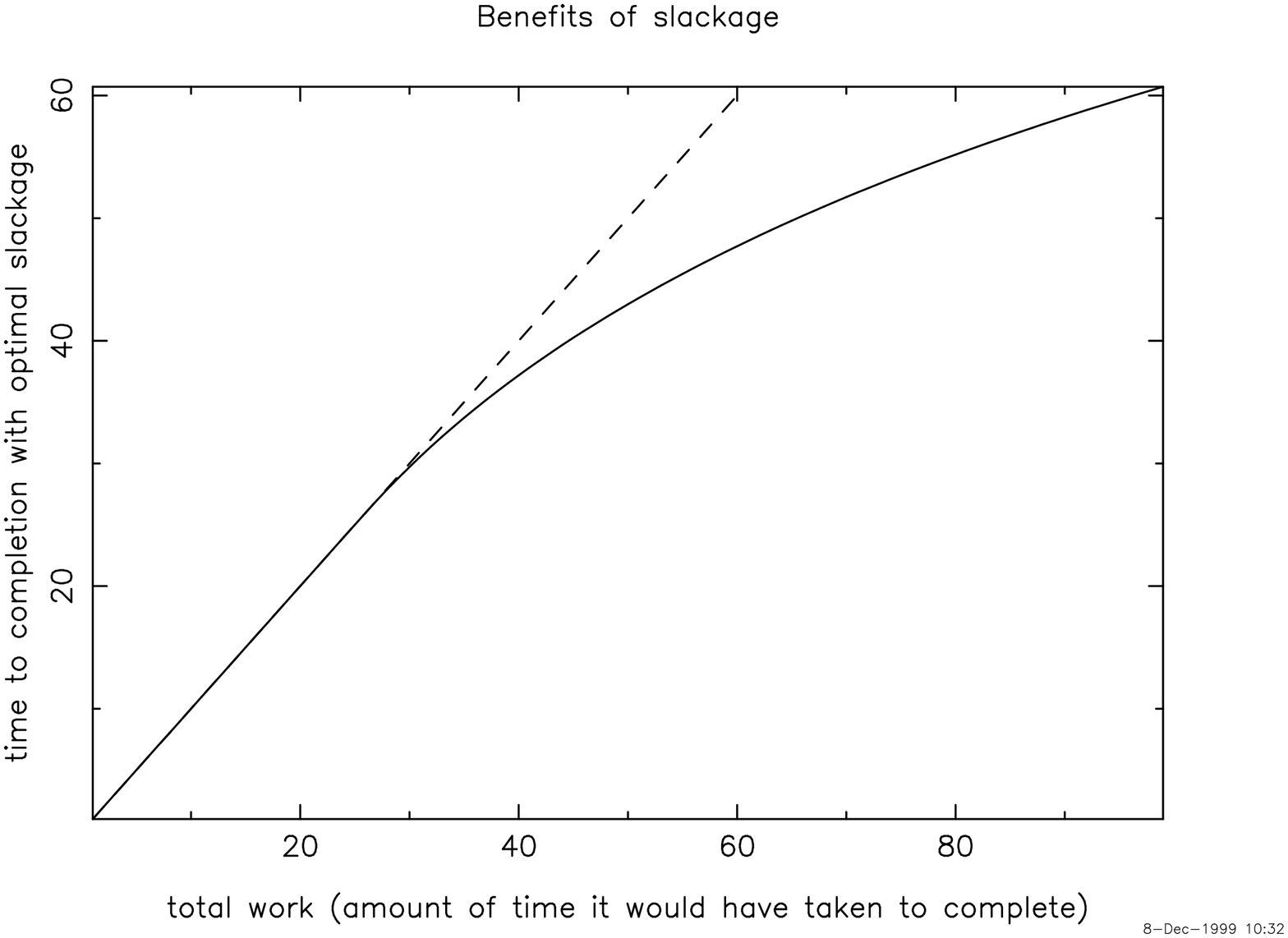}}}
\end{figure}

\vspace{5mm}

The conclusion that is drawn from this is that simulations done with a 
$t_c$ (26 month) runtime on the best machine that can be purchased for a given 
cost when the calculation begins are indeed optimal in the sense of 
utilization of ever improving computer resources. They are the largest 
calculations that should be done with the present resources. The optimal 
time to begin any more arduous computation is in the future (after an 
optimal amount of slack time). Furthermore any more trivial calculation 
should have been started in the past because calculations smaller than 
$t_c$ months runtime complete in the order they are undertaken.
\footnote{You may 
notice that this regime corresponds to a negative $s\star$ parameter, however 
we choose to neglect this notion since it requires postulating the possibility of anti-slack.}

This suggests that for any given calculation there is a best time
to start, and that a valid strategy would be to always attempt
problems that optimally utilize the resources. Obviously the effect
of Moore's law is that that the optimal problem scales as the
rate of computation. Our calculations place a normalization on this 
scale and suggest that you will get the best possible performance if you 
choose to attack problems that will take $t_c$ months to run on your computer
when you get around to starting the computation.

\bigskip
\noindent \textbf{References}

\medskip
\noindent Moore, G.E. 1965, \emph{Electronics} (Volume 38 Number 8), pp. 114--117

\end{document}